\documentstyle[]{aa}
\input epsfig.sty

\newcommand{\be}{\begin{equation}}
\newcommand{\ee}{\end{equation}}
\newcommand{\bea}{\begin{eqnarray}}
\newcommand{\eea}{\end{eqnarray}}
\def\divop{\mathop{\rm div}\nolimits}
\newcommand{\od}{{\rm d}\kern.05em}
\def\odf#1#2{\frac{{{\rm d}\kern.05em}#1}{{{\rm d}\kern.05em}#2}}
\begin{document}
 \title{Magnetic amplification in cylindrical cosmological structure}  

\author{Gra\.zyna Siemieniec-Ozi\c{e}b{\l}o \and Zdzis{\l}aw A. Golda}


\institute{Astronomical Observatory, Jagellonian University\\
Faculty of Physics, Astronomy and Applied Computer Science\\
ul. Orla 171, 30--244 Krak\'ow, Poland}
\titlerunning{Magnetic amplification in cylindrical cosmological structure}
\authorrunning{G. Siemieniec-Ozi\c{e}b{\l}o and Z. A. Golda}

\abstract{   
We derive the amplification of the cosmological magnetic field
associated with forming gravitational structure. The
self-similar solutions of magnetohydrodynamic equations are computed both in 
linear and nonlinear regimes. We find that the relatively fast
magnetic field enhancement becomes substantial in the nonlinear
phase.}

 \maketitle  


\section{Introduction}
\label{sec:wstep}

The hypothesis describing the dynamical role  of the primordial
magnetic field in the formation and evolution of gravitational
structure frequently occurs in the literature (e.g.~Wasserman 
1978; Kim~et~al. 1996; Peebles 1995). On the contrary, the
inverse trend i.e. the amplification of the magnetic field during 
density perturbation collapse is not often represented in
the context of large-scale structure formation. Common
practice is to refer to the constraints for magnetic field
amplification set by the density of collapsed matter
(e.g.~Zeldovich et al. 1980).  The hints of magnetic field
existence on cosmological scales excites interest not only in
the absolute value of a frozen field magnification but also in
its growth rate. The early nonlinear and previrial phase of the
gravitational formation is of particular importance, since it
results in several megaparsec structures observed as
superclusters or filaments. An understanding of the 
amplification rate of the structure is obviously related to the
explanation of the appearance of sufficiently strong frozen-in
magnetic fields expected at this stage of collapse.

We investigate the mildly nonlinear collapse of cylindrical
gravitational structure and give the growth 
rate of the primordial magnetic field as a function of the accretion 
velocity field. The magnetic field growth proceeds intensively
during the phase of fluid compression. The magnetic flux for
collapsing plasma is conserved and the magnetic strength changes
according to the induction equation. It clearly shows that 
substantial amplification occurs for strongly compressing flows
i.e. for growing $\divop \textbf{v}$ --- analogously to the
shock processes. The general expression for $\divop \textbf{v}$
is obtained thanks to the self-similar form of the hydrodynamic
equation. The self-similar presentation of 
magnetohydrodynamics becomes possible in the case of rapid
density contrast and velocity evolution, when $v$ and $\delta
\propto a^n$ $(n > 1)$ i.e.~when the Lorentz force neglect in the Euler
equation is justified.

The plan of the paper is as follows. In Section~\ref{sec:podstawowe_rownania} 
we give the magnetohydrodynamic
(MHD) equations for cylindrical structures in comoving 
coordinates of a flat universe. We also discuss the assumptions
and symmetries allowing us to separate the induction equation. In
Section~\ref{sec:cylindryczne_perturbacje} we derive the
self-similar set of hydrodynamic equations. Its linearization
and the subsequent comparison with the known 
velocity solutions are also given. We obtain the general
nonlinear relation $\divop \textbf{v}$ versus $\delta$. The rate
of magnetic field amplification is discussed in 
Section~\ref{sec:nieliniowe_wzmocnienie} both in the linear and nonlinear
regime. In the former we present the analytical expression for
amplification and the results of numerical integration in the
latter.

\section{Basic equations and perturbation scheme for magnetized fluid}
\label{sec:podstawowe_rownania}

The magnetic field modifies the standard equations for density
perturbations and peculiar velocities. These equations completed
by the Faraday equation for the magnetic field have been extensively
studied in the context of magnetic field influence on galaxy
formation (e.g. Lesch \&~Chiba 1995). The linear approximation of the 
MHD equations has been analyzed in many papers.
In the extreme approach, the primordial magnetic field is 
responsible for the origin of density fluctuations (Wasserman 
1978). In his paper the compression of matter is induced by the
Lorentz force but the linearized hydrodynamic equations keep
both $\rho$ and {\bf v} as small linear values. There is no
fluid back reaction on the magnetic field.

In this paper we apply a different approach. The weak initial
and uniform magnetic field {\bf B} is embedded in the forming
large-scale structure which evolves in the dust era. The
gravitational collapse changes the strength of the magnetic field in 
the linear and then nonlinear regime. We are interested in the
nonlinear amplification rate of the ordered field.

The MHD equations are in comoving coordinates given by (e.g.
Wassermann 1978) 
	\bea
{\partial\rho \over \partial t} + 3 {\dot a \over a}{\bf\nabla\cdot v} + 
{1 \over a}{\bf\nabla\cdot}(\rho{\bf v}) &=& 0,\label{eq:01}\\
{\partial {\bf v} \over \partial t} + {\dot a \over a}{\bf v} + 
{1 \over a}({\bf v {\cdot} \nabla}) {\bf v} &=&-{1 \over a} {\bf \nabla}\psi  -
{1 \over 4 \pi a \rho} {{\bf B}{\times} (\nabla {\times} {\bf B}}),\nonumber\\
&&{}\label{eq:02}\\
{1 \over a^2}\triangle \psi &=& 4 \pi G (\rho - \rho_0),\label{eq:03}\\
{\partial {\bf B} \over \partial t} + 2 {\dot a \over a} {\bf B} &=&
{1 \over a}{\bf \nabla \times} ({{\bf v} \times {\bf B}}),\label{eq:04}\\
{\bf \nabla \cdot B} &=& 0,\label{eq:05}
	\eea
where $\rho(t, \bf{r})$ is the matter density
($\rho_0(t)$ means the background density), ${\bf v}(t, \bf{r})$ is the
peculiar velocity of the matter forming the structure and $\psi
(t, \bf{r})$ is the density perturbation potential; $a(t)$ is the
scale factor given in units $a(t_0)$. For an assumed flat FRW
cosmological model we take the initial value of the uniform magnetic
field $B_0$ at the initial time~$t_0$.

The temporal evolution analyzed here is restricted to the after
recombination phase and we do not discuss the magnetogenesis
processes assumed to occur prior to this phase. The small value
of the initial magnetic energy density (relative to the matter
density) is assumed. Since the field uniformity is maintained
within the scale of the forming structure, it does not perturb
the background isotropy. The evolving structure and thus the
considered field have axial symmetry. The velocity of the 
accreting fluid is taken for simplicity as ${\bf v} =
[v_r(t, r), v_{\theta}(t, r), 0]$.  The assumed cylindrical
symmetry is consistent with the isotropy and homogeneity of the
FRW model for all length scales large with respect to the
coherence scale.  Finally, the nonlinear perturbation equations
for density fluctuations of presureless matter, velocity
field and the magnetic field $\bf B$ are given by
	\bea
{\partial\delta \over \partial t} + {1 \over a}{\bf\nabla\cdot v} + 
{1 \over a}{\bf\nabla\cdot}(\delta{\bf v}) &=& 0,\label{eq:06}\\
{\partial {\bf v} \over \partial t} + {\dot a \over a}{\bf v} + 
{1 \over a}({\bf v \cdot \nabla}) {\bf v} &=& 
-{1 \over a} {\bf \nabla} \psi
-{1 \over 4 \pi a \rho} ({\bf B_0 \times \nabla \times b}\nonumber\\
&&{}+
{\bf b \times \nabla \times b}),\label{eq:07}\\
{1 \over a^2}\triangle \psi &=& 4 \pi G (\rho - \rho_0),\label{eq:08}\\
{\partial {\bf B} \over \partial t} + 2 {\dot a \over a} {\bf B} &=&
{1 \over a}{\bf \nabla \times} ({\bf v \times B}),\label{eq:09}
	\eea
where $\delta(t, r) = (\rho - \rho_0)/ \rho_0(t)$ is the matter
density contrast and ${\bf b}(t, r)$ is the magnetic field
perturbation i.e. ${\bf B}(t, r) = {\bf B_0}(t) + {\bf b}(t, r)$\footnote{ 
For simplicity, it is assumed here that the background 
field is represented by its uniform component. Following Barrow et al. 
(1997), its sufficiently small initial value (i.e. $B_0 < 10^{-9}$ G) 
does not affect either the background dynamics or the CMB spectrum. 
The obtained final results on B distribution are not expected to depend on 
this (unrealistic on a few Mpc scale) assumption.}.
In the
epoch when the condition of low energy density associated
with the magnetic field is valid i.e. $B^2/ \rho  \ll 1$, the
evolutionary perturbation scheme may be simplified. The last
expression in the Euler equation consists of two terms; the first 
may be ignored due to the small value of $\bf B_0$ and the
second is proportional to $\nabla \rho \propto \nabla |{\bf b}|$, 
 as can be seen from eqs. (\ref{eq:11}) and (\ref{eq:20}) below.  We shall
assume in the first approximation that it is also small compared
to the density potential gradient. Thus in the first iteration
step we postulate a force-free magnetic field. The Faraday
equation describes the magnetic field scaling as a function of
accretion velocity. Both the amplified magnetic field and the 
density compression are coupled with the velocity field but the
induction equation decouples from the hydrodynamic equations.
Therefore its formal solution may be given independently.
Applying the assumed symmetries and transforming the r.h.s. of 
eq.~(\ref{eq:09}) one obtains the induction equation in the integrable form.
	\bea
\odf{{\bf B} a^2}{t} \equiv {\partial\, {\bf B} a^2 \over \partial t}
+ \left({{\bf v} \over a}{{\cdot} \nabla}\right)\left({\bf B} a^2\right)=
-\left({\bf B} a^2\right) \nabla {\cdot}\left({{\bf v} \over a}\right).\nonumber\\
\label{eq:10}
	\eea
Its formal solution is
	\bea
 {\bf B} a^2 = {\bf B_0} \exp\left[{-\int{\bf \nabla \cdot}\left({{\bf v} \over a}\right) \od t}\right] ,
 \label{eq:11}
	\eea
where $\bf v$ as well as ${\rho}$ and ${\psi}$ are given by the
solutions of eqs.~(\ref{eq:06}--\ref{eq:08}). The successive iterative procedure
will couple in the next steps the Faraday and gravitational
perturbations equations, including the magnetic force to the r.h.s.
of Euler equation. However this is not essential before $|{\bf B}|
\approx \psi$.

\section{Nonlinear cylindrical perturbation}
\label{sec:cylindryczne_perturbacje}

The dynamics of collapsing structure modeled from a 
pressureless fluid is governed by eqs.~(\ref{eq:06}--\ref{eq:08}). Applying the
assumed form of accretion velocity, the dynamic equations for
the fluid transform the continuity equation into
	\bea
{\partial \delta \over \partial t} + {1 \over ar} {\partial\,
rv_r \over \partial r} +  {1 \over a} \left ({\delta \over r}
{\partial (rv_r) \over \partial r} + v_r {\partial \delta \over
\partial r} \right ) = 0, 
\label{eq:12}
	\eea
two components of the Euler equation become
	\bea
{\partial v_r \over \partial t} + {\dot a \over a}v_r + {1 \over a} 
\left ({v_r \over r} {\partial\, rv_r \over \partial r}
- {1 \over r^3} (r^2 v_r^2 + r^2 v_{\theta}^2) \right ) &=& 
-{1 \over a} {\partial \psi \over \partial r},\nonumber\\
&&\label{eq:13}\\
{\partial v_{\theta} \over \partial t} + {\dot a \over
a}v_{\theta} + {1 \over a}  {v_r \over r} {\partial\,
rv_{\theta} \over \partial r} &=& 0\label{eq:14}. 
	\eea
The peculiar gravitational potential satisfies the Poisson equation
	\bea
{1 \over ra^2} {\partial \over \partial r} \left (r{\partial \psi \over
\partial r} \right ) = \alpha \delta
\label{eq:15} 
	\eea
where $\alpha (t) \equiv 4 \pi G \rho_0$.\pagebreak[4]

Many authors (see for example, Lifshitz 1946; Groth \& Peebles 1975; 
Peebles 1980) have studied  the
analytical solutions for the linearized problem of perturbation
evolution in an expanding Universe. In several cases the nonlinear
solutions have also been developed (Zeldovich 1970). Yet, in
general, the high degree of complexity of the nonlinear set of
partial differential equations does not allow one to obtain 
analytical solutions. In this paper we also attempt to obtain
the particular nonlinear solution for growing structure.
Obviously, to proceed, additional assumptions are necessary.
The problem becomes much simplified when we adopt that the
tangential velocity component is small relative to the radial
one i.e.  $(v_{\theta}/v_r)^2 \ll 1$.  This results in a 
separation of eq.~(\ref{eq:14}). The above assumption is consistent with
the small value of the initial vorticity and the resulting
equations reduce to the equation set (\ref{eq:12}--\ref{eq:13}) and (\ref{eq:15}), 
describing the rotationless flow of the magnetized fluid.

To obtain a convenient form, we choose first the new time
coordinate  $x =  a^{-1/4} =  t^{-1/6}$, instead of the cosmic
time $t$. Our analysis becomes more clear when we use the
rescaled variables.  Replacing the old functions $\delta , v_r, 
\psi$  by the new variables $X, z_2, z_4$,  where $X = \delta,
z_2 = a r v_r$ and $z_4 = r^3 a^2 {\partial \psi \over \partial
r}$ and defining the new independent variable $\xi \equiv r/x$,
the continuity, the Euler and Poisson equations read
	\bea
z_2'(\xi)(1+X(\xi))+\left(\frac{1}{6} \xi^2+z_2(\xi)\right)X'(\xi)&=&0,
\label{eq:16}\\
-z_2^2(\xi)+\xi\left(\frac{1}{6} \xi^2+z_2(\xi)\right) z_2'(\xi)+z_4(\xi)&=&0,
\label{eq:17}\\
-\beta \xi^4X(\xi)-2z_4(\xi)+\xi z_4'(\xi)&=&0,
	\label{eq:18}
	\eea
where prime means the differentiation with respect to $\xi$ and
$\beta  \equiv 4 \pi G (\rho_0 a^3)$ 
(hereafter we take its value as $\beta = \frac{2}{3}$).  This scaling is not only 
a mathematical manipulation. Introducing the $\xi$ ---
variable reduces the differential complexity of the problem,
allowing us to substitute the partial by the ordinary differential
equations. In effect, the equations (\ref{eq:16}--\ref{eq:18}) govern the
self-similar behaviour of the fluid motion, leading to the
nonlinear 3$^{\rm rd}$--order equation for the rescaled velocity $z_2(\xi)$
	\bea
&&\xi^3\left[\frac{1}{6} \xi^2 + z_2(\xi)\right]^2
z_2^{(3)}(\xi)+\xi^2\left[\frac{1}{6}\xi^2+z_2(\xi)\right]\nonumber\\
&&{}\cdot\left[-5z_2(\xi)+\xi\left(\frac{1}{6}\xi+4z_2'(\xi)\right)\right]z_2''(\xi)\nonumber\\
&&{}-\xi\left[-15 z_2^2(\xi)+4\xi z_2(\xi)\left(-\frac{1}{2}\xi + 2z_2'(\xi)\right)\right.\nonumber\\
&&\left.+\xi^2\left(\frac{25}{36} \xi^2+\frac{2}{3}\xi z_2'(\xi)-z_2'^2(\xi)\right)\right]z_2'(\xi)\nonumber\\
&&-8\left[\frac{1}{6}\xi^2 +z_2(\xi)\right]z_2^2(\xi)=0.
	\label{eq:3rzedu}
	\eea

The detailed analysis of the above dynamical set \mbox{(\ref{eq:16}--\ref{eq:18})} may
be performed either through the qualitative methods of its
asymptotic form or requires numerical integration.

Nevertheless, some general relation may be derived even before
the solutions are obtained. The continuity equation (\ref{eq:16}) implies
the generalized expression for density contrast and  velocity
function $z_2$.
	\bea
\ln (1 +X) = -\int{z_2^\prime(\xi)\over \frac{1}{6} \xi^2 + z_2(\xi)} \od\xi 
 = - \int {{\bf \nabla \cdot v} \over a}\od t.
	 \label{eq:20}
	\eea

This is a nonlinear counterpart of the known linear relation
i.e. $\divop {\bf v} \propto X$ (e.g. Willick \& Strauss 1998), coupling $\rho$ vs $v$ and
thus $B$ vs $v$.  Lacking the analytical solutions we note 
that the asymptotic form for large velocities $|z_2| \gg  \xi
^2$ predicts the asymptotic property of the density contrast $ X
+ 1 \cong |z_2|^{-1}$.

The linearization of the eqs. (\ref{eq:16}--\ref{eq:18}) and its resulting
solution for $z_2$ allows us to compare the time dependent part of
$v_r$ with the linear regime expressions $v_{1, 2}
\propto a^{1/2}, a^{-2}$ (Peebles 1980). Finally, one has
	\bea
(z_2^{\rm lin})_{1}  = - \xi^6,~~(z_2^{\rm lin})_{2}  = - \xi^{-4},
	\label{eq:21a}
	\eea
and the implied linear solutions for density contrast are
	\bea
(X^{\rm lin})_{1}  = 9 \xi^4,~~(X^{\rm lin})_{2}  = 4 \xi^{-6}.
	\label{eq:21b}
	\eea

 The analysis of the asymptotic behaviour of eqs. (\ref{eq:16}--\ref{eq:18}) 
at infinity ($\xi\to\infty$) becomes straightforward when we define the new  
variable $\zeta$, $\xi\to 1\slash\zeta$. The subsequent linearization of the above 
equation set gives for $\xi\to\infty$ the following solutions for the 
velocity measure 
\bea
(z_2^{\rm lin})_{1}  = - \zeta^{-6},~~(z_2^{\rm lin})_{2}  = - \zeta^{4},~~(z_2^{\rm lin})_{3}  = \mbox{const},
	\label{eq:21aa}
	\eea
and the density contrast 
	\bea
(X^{\rm lin})_{1}  = 9 \zeta^{-4},~~(X^{\rm lin})_{2}  = 4 \zeta^{6}.
	\label{eq:21bb}
	\eea
The above enables us to state that both velocity field and density 
contrast have solutions vanishing for $\xi\to\infty$. 
Below we deal with the physical solutions 
i.e. regular at $\xi = 0$ and at $\xi\to\infty$.

\section{Linear and nonlinear magnetic amplification}
\label{sec:nieliniowe_wzmocnienie}

The adiabatic matter compression determines the field
amplification through the reached value of $\rho$.  The
amplification degree is controlled by the compression symmetry
(see eq.~(\ref{eq:10})). For cylindrically collapsing structures one obtains
the magnetic field amplification $W_B \equiv {B \over B_0} a^2$, 
thus taking into account eqs.~(\ref{eq:11}) and (\ref{eq:20}) 
one has ${W_B \over 1+X} =\mbox{const}$, while
for example, in the case of spherical symmetry this quantity
would be given by  ${W_B \over (1+X) ^{2/3}} = \mbox{const}$.
On the other hand, the frozen field amplification expressed by
the solution of the induction equation is defined by the velocity
field given by eq.~(\ref{eq:3rzedu}). To represent it directly by the time
evolution of the velocity field we first present the velocity
divergence in old variables  ($x, r$)
	\bea
{\bf \nabla \cdot v} = {1 \over r} {\partial(rv_r) \over
\partial r} = \xi^{-1} x^2 {dz_2 \over d\xi} = -{x^5 \over r^2} 
{\partial z_2 \over \partial x}.
	 \label{eq:22}
	\eea
Note that the `self-similar' variable $\xi$ is used here only
for differentiation, since we have $x {\partial 
\over \partial r} = {\od  \over \od\xi} = - {x \over
\xi}{\partial  \over \partial x}$.  The exponent in eq.~(\ref{eq:11})
becomes
	\bea
\int {{\bf \nabla \cdot v} \over a} \od t = {1 \over  r^2} \int x^2
{\partial z_2 \over \partial x} \od x.
	 \label{eq:23}
	\eea
Replacing, in the above, the time variable by the scale factor  $a$
and assuming the separability of the velocity function  $v_r = -
V(r) v(a)$ (for collapse we have $v_r (r,t) < 0$ ), one obtains
the alternative form of eq.~(\ref{eq:23}).
\bea
-\int {{\bf \nabla \cdot v} \over a} \od t = {V(r) \over r}
\int \left ( a^{1 \over 2} {\partial v(a) \over \partial a} +
v(a) a^{-\frac{1}{2}} \right ) \od a.
	 \label{eq:24}
	\eea
Thus the final expression for density and magnetic field
compression due to nonlinear structure formation reads
	\bea
W_B &=& X+ 1 = \exp \left[{V(r) \over r}
\int \left[a^{1 \over 2} {\partial v(a) \over \partial a} +
v(a)a^{-\frac{1}{2}} \right]\od a \right].\nonumber\\
	 &&{}\label{eq:25}
	\eea 
Combining the above equation, for the growing mode ($v(a)
\propto a^{1/2}$) of the velocity field and eq.~(\ref{eq:21a}) yields the
following amplification rate in the linear regime
	\bea
W_B = \exp \left[{3\, V(r) a \over 2\, r}\right]
 = \exp \left[{\frac{3}{2}\, r^4 a }\right] = \exp \left[{\frac{3}{2}\, \xi ^4
}\right].
	 \label{eq:26}
	\eea 
The successive discussion of nonlinear behaviour will 
also deal with the growing mode. Knowledge of the exact nonlinear amplification $W_B =
\exp[{-\int{\bf \nabla \cdot} ( {{\bf v} \over a})}\od t] = 
\exp[-\int{z_2^\prime(\xi)\over \frac{1}{6} \xi^2 + z_2(\xi)} \od\xi ]$ needs
the solution of the very complex, quasilinear equation (\ref{eq:3rzedu}) for $z_2$.
 The zero value of the coefficient at the $z_2^{(3)}$ term allows us  
to determine the singularity points of this equation i.e. $\xi=0$ 
and $\frac{1}{6}\xi^2+z_2=0$. 
Since we postulate the regular velocity solution at the structure centre 
we focus here on a later one, which appears when the overdensity  
grows in the nonlinear regime. The singularity point 
position depends on the chosen boundary condition (a movable singularity). 
This is specific to the nonlinear differential equations ~(Bender, \&~Orszag 1978). 
Physically, the emergence of a singularity in the velocity and density field is 
an artefact of the collisionless fluid. This feature has also been found  
in $N$-body simulations and the test-particle approach (e.g. Bertschinger, 1985) 
to gravitational collapse, as an outer caustic which forms around the structure. 
(For collisional matter, $p \neq 0$, it becomes a shock surface.)

Since its analytical solution  is difficult to obtain, below we
present the results of numerical integration below the caustic surface.  
Independently we verify the asymptotic behaviour of the solution
at  $\xi \rightarrow \infty$. As expected from eq.~(\ref{eq:21aa}), 
the velocity function $z_2(\xi)$ and the density contrast $X(\xi)$ approach zero at infinity.  
The principal features of the particular solution inside the structure 
and its asymptotic behaviour at infinity are 
illustrated in Fig.~\ref{fig:01} and Fig.~\ref{fig:02} below. These figures show
the density contrast and the velocity field as a function of $\xi$ (or~$\zeta$) in 
the log-log scale. Both quantities are 
expressed respectively, in units of $(X^{\rm lin})_1$ and $(z_2^{\rm lin})_1$  
or in units of $(X^{\rm lin})_2$ and $(z_2^{\rm lin})_2$ at infinity.
 \begin{figure}[h]  
\centerline{\psfig{file=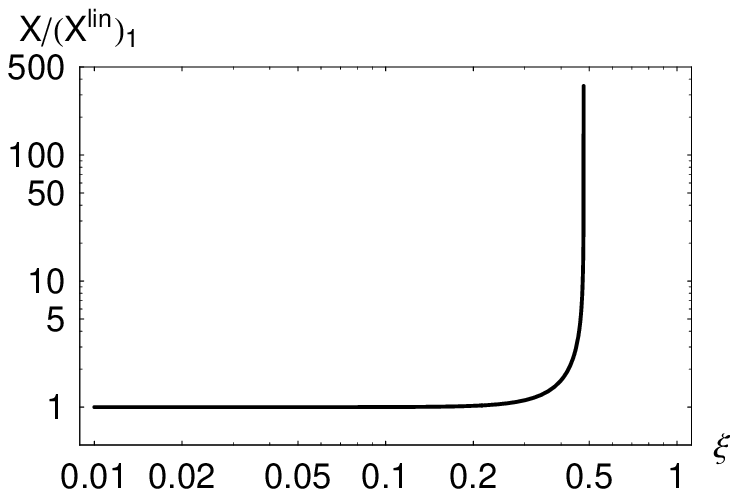,width=0.25\textwidth}\hspace{-0.5em}\psfig{file=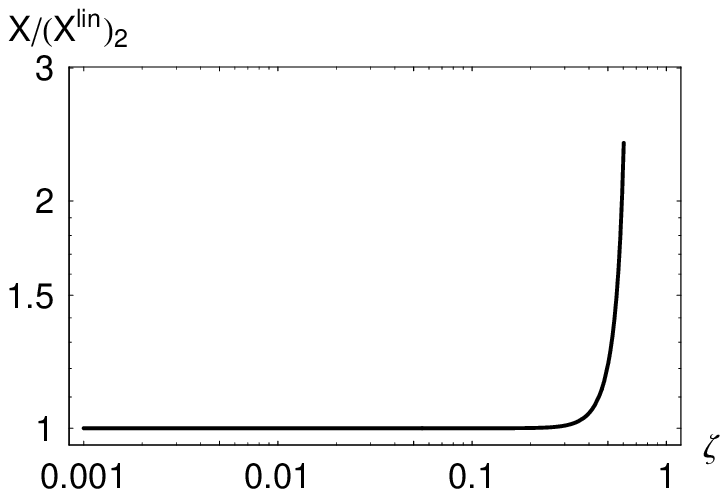,width=0.25\textwidth}}
\caption{The nonlinear excess of the density contrast $X/X^{\rm lin}$ below the discontinuity (left panel),  
as a function of the \mbox{$\xi$-variable} and (right panel), above the discontinuity  
 as a function of the $\zeta$-variable.}
     \label{fig:01}  
     \end{figure}
\begin{figure}[h]
\centerline{\psfig{file=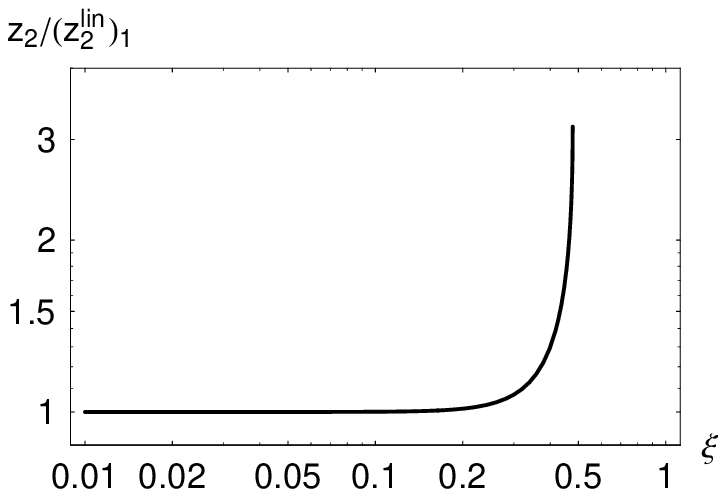,width=0.25\textwidth}\hspace{-0.5em}\psfig{file=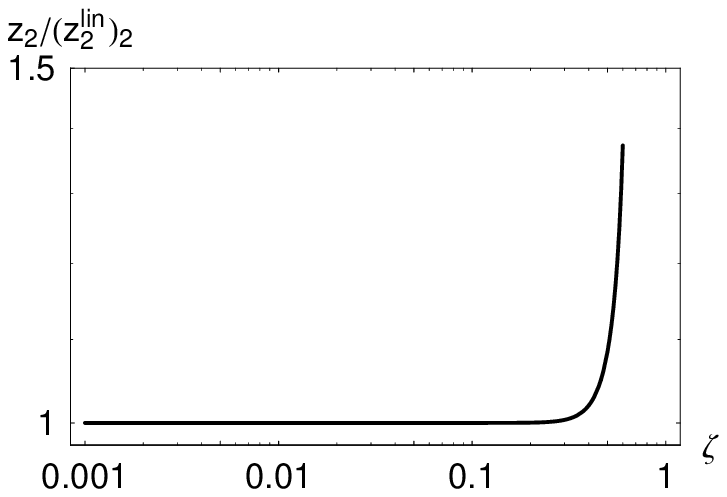,width=0.25\textwidth}}
\caption{The nonlinear excess of the velocity measure $z_2/z_2^{\rm lin}$ below the discontinuity (left panel),  
as a function of the \mbox{$\xi$-variable} and (right panel), above the discontinuity  
 as a function of the $\zeta$-variable.}
\label{fig:02}
	\end{figure}	
The presented profiles are only qualitative, in the sense that
they are given by the numerical integration for a required set of the particular boundary conditions. 
The boundary conditions for numerical integration were taken 
from the linear solutions of eqs. (\ref{eq:16}--\ref{eq:18}).  
They clearly display two distinctive regions corresponding
to linear and nonlinear formation rates. Roughly, the linearity  
is represented by the straight line $\frac{X}{(X^{\rm lin})_1} \cong 1$ 
and the nonlinear region of the magnetic field and density contrast 
amplification for $\xi \geq 0.2$ shows a strongly 
growing curve. It should be noted that the validity of the linear approach 
is slightly different in the case of density contrast and the 
velocity function\footnote{The density growth $\frac{X}{(X^{\rm lin})_1}$ 
is faster than the velocity growth $\frac{z_2}{(z_2^{\rm lin})_1}$. 
It is clearly seen when the respective values are compared i.e. 
$\frac{X}{(X^{\rm lin})_1}(0.2)=1.02$, ~$\frac{X}{(X^{\rm lin})_1}(0.47)=6.15$ 
and $\frac{z_2}{(z_2^{\rm lin})_1}(0.2)=1.01$, 
~$\frac{z_2}{(z_2^{\rm lin})_1}(0.47)=2.18$.}.

The direct comparison of the linear and nonlinear
amplification rate for a given structure radius shows that the nonlinear
compression implies the magnetic field growth rate of several
orders of magnitude. The predicted (cf.~Fig.~\ref{fig:01}) amplification
of the order of $W_B \sim 10^3$  within the range ($\xi_1,
\xi_2$) = (0.28, 0.48) corresponds to the evolutionary period given by
$\frac{a_2}{a_1} = (\frac{\xi_2}{\xi1})^4 \approx 6.5$.  Thus,
the magnetic growth of the order of $2\times 10^3$ may be attributed to the
early nonlinear structure evolution between, for instance,
redshift $\sim 6$ and  redshift  $\sim 0$. Certainly, the magnetic 
amplification becomes much greater as nonlinear evolution proceeds 
but then the validity of the applied assumptions becomes questionable.

\section{Conclusions}

The goal of this paper is to present the amplification rate of the magnetic field
associated with the forming gravitational structure of cylindrical symmetry. 
The widespread conviction that the large-scale structures are filled with  
microgauss cosmological magnetic fields motivates our interest in amplification processes 
during their evolution. On the other hand, we see a high degree of filamentarity 
 in the galaxy redshift surveys (e.g. Sathyaprakash et al. 1998). 
This demonstrates that we deal with magnetized, 
elongated structures of axial symmetry. 

To determine their 
magnetic structure growth, several simplifications are needed. We used here two 
categories of simplifying assumptions: physical (i.e. $p \sim 0$ and $F_L \sim 0 $), constraining 
the results to the early nonlinear phase and geometrical ones --- requiring the radial motions 
and thus the axial fields. On the basis of a formal solution of the induction equation we 
obtained the exact analytical expression for linear field amplification. The relationship 
between the density contrast and the magnetic field strength is established through 
the velocity field divergence. However the major conclusion concerns the nonlinear 
phase. The magnetic field may be effectively enhanced there. The density contrast growth 
is stronger than the velocity, achieving the nonlinearity regime earlier. 
The radial structure of the magnetic field and density contrast are identical, in general 
--- nonhomogeneous.  

 Contrary to this highly idealized model, 
in the realistic situation, the centrifugal forces will stop the collapse. 
This will however occur in the successive, virialization phase, when the 
matter will become collisional and then shocked. Therefore, a proper description requires  
more elaborate application of the fluid model. Within its current limitations 
the above applied symmetries seem to be less weighty than the physical assumptions. 
According to previous papers (e.g. Siemieniec \& Woszczyna 2004, Bruni et al. 2003) 
the more degenerate, pancake geometry 
leads to comparable amplification results. Introducing 
cylindrical symmetry enables us instead to depict the magnetic field profile 
inside the structure. The substantial enhancement of the matter density 
accreting onto collapsing structure indicates that significant magnetic 
fields may be produced in its outer region --- the precursor of the future 
shock.

\section*{Acknowledgements} 
  
We thank the referee, Dr. G. Sigl for criticisms and useful comments. 
This work was supported in part by `Komitet Bada\'n Naukowych' projects: 
`Cosmic magnetic fields' (G.~S-O) and `Rigorous description of scalar and tensor perturbations 
in flat and open Friedmann universes' (Z.A.~G).

\def \A&A{{A\&A}}
\def \ApJ{{ApJ}}
\def \ApJS{{ApJS}}
\def \JP{{J. Phys. \rm(USSR)}}

\section*{References}   

\parskip=0pt   
\parindent=7mm
\noindent
Barrow, J., Ferreira, P., \& Silk, J., 1997, PRL, 78, 3610\\
Bender,~C.M.,~\&~Orszag,~S.A.~1978,~Advanced~Mathemat\-ical Methods for Scientists and Engineers 
(McGraw-Hill Book Company, New York)\\
Bertschinger, E., 1985, \ApJS, 58, 39 \\
Bruni, M., Maartens, R., \& Tsagas, C., 2003, MNRAS, 338, 785\\
Groth, E., \& Peebles, P., 1975, \A&A, 41, 143\\
Kim, E., Olinto, A., \& Rosner, R., 1996, \ApJ, 468, 28\\
Lesch, H., \& Chiba, M., 1995, \A&A, 297, 305\\
Lifshitz, E., 1946, \JP, 10, 116\\
Peebles, P., 1980, The Large-Scale Structure of the Universe (Princeton Univ. Press, Princeton)\\
Peebles, P., 1993, Principles of Physical Cosmology (Princeton Univ. Press, Princeton)\\
Sathyaprakash, B., Sahni., \& Shandarin, S., 1998, \ApJ, 508, 551\\
Siemieniec, G., \& Woszczyna, A., 2004, \A&A, 414, 1 \\
Wasserman, I., 1978, \ApJ, 224, 337\\
Willick, J., \& Strauss, M., 1998, \ApJ, 507, 64 \\
Zeldovich, Y., 1970, \A&A, 5, 84\\
Zeldovich, Y., Ruzmaikin, A., \& Sokolov, D., 1980, Magnetic Fields in Astrophysics (McGraw-Hill, NY)

\end{document}